# THE EVOLUTION OF INSURANCE PURCHASING BEHAVIOR: AN EMPIRICAL STUDY ON THE ADOPTION OF ONLINE CHANNELS IN POLAND

Gabriela WOJAK[1], Ernest GÓRKA[2], Michał ĆWIĄKAŁA[3], Dariusz BARAN[4], Rafał ŚWINIARSKI[5], Katarzyna OLSZYŃSKA[6], Piotr MRZYGŁÓD[7], Maciej FRASUNKIEWICZ[8], Piotr RĘCZAJSKI[9], Daniel ZAWADZKI[10], Jan PIWNIK[11]

[1] Pomorska Szkoła Wyższa w Starogardzie Gdańskim, Instytut Zarządzania, Ekonomii i Logistyki; gabriela.wojak@twojestudia.pl, ORCID: 0009-0003-2958-365X
[2] Wyższa Szkoła Kształcenia Zawodowego; ernest.gorka@wskz.pl, ORCID: 0009-0006-3293-5670
[3] Wyższa Szkoła Kształcenia Zawodowego; michal.cwiakala@wskz.pl, ORCID: 0000-0001-9706-864X
[4] Pomorska Szkoła Wyższa w Starogardzie Gdańskim, Instytut Zarządzania, Ekonomii i Logistyki; dariusz.baran@twojestudia.pl, ORCID: 0009-0006-8697-5459
[5] I'M Brand Institute sp. z o.o.; r.swiniarski@imbrandinstitute.pl, ORCID: 0009-0005-6226-943X
[6] Polsko Japońska Akademia Technik Komputerowych; kontakt@olszynska.com, ORCID: 0009-0003-4309-6233
[7] Piotr Mrzygłód Sprzedaż-Marketing-Consulting; piotr@marketing-sprzedaz.pl, ORCID: 0009-0006-5269-0359
[8] F3-TFS sp. z o.o.; m.frasunkiewicz@imbrandinstitute.pl, ORCID: 0009-0006-6079-4924
[9] MAMASTUDIO Pawlik, Ręczajski, spółka jawna; piotr@mamastudio.pl, ORCID: 0009-0000-4745-5940
[10] GLOBAL HYDROGEN spółka akcyjna; daniel.zawadzki@globalhydrogen.pl, ORCID: 0009-0001-4783-3240
[11] WSB Merito University in Gdańsk, Faculty of Computer Science and New Technologies; jpiwnik@wsb.gda.pl, ORCID: 0000-0001-9436-7142

**Purpose:** The aim of this study is to examine the evolution of insurance purchasing behavior in Poland, with a focus on the growing role of online distribution channels. The paper seeks to understand the motivations, preferences, and concerns influencing consumer choice between digital and traditional methods.

**Design/methodology/approach**: The research uses a quantitative approach based on a structured online survey of 100 respondents. It explores demographics, purchasing habits, channel preferences, and decision-making factors such as price, trust, and convenience. The analysis identifies behavioral patterns related to digital adoption.

**Findings:** The results show a clear increase in the use and acceptance of online channels, especially for simple insurance products. Consumers appreciate the speed, availability, and price comparison offered by digital tools. However, some still prefer traditional channels for complex purchases due to trust and clarity. A hybrid approach is emerging, combining digital access with agent support.

**Research limitations/implications**: The study is limited by its sample size and focus on the Polish market. It does not explore differences by insurance product type. Future research could expand the sample and analyze specific product categories or the role of AI tools in shaping digital insurance experiences.

**Practical implications:** The findings suggest that insurers should improve the usability, transparency, and security of digital platforms. At the same time, maintaining advisory services is essential to support customers who seek personal guidance. Omnichannel strategies appear to be the most effective.

**Social implications:** Digital channels enhance access to insurance, especially for tech-savvy or remote users. Addressing concerns about trust and data privacy may increase consumer confidence and promote digital inclusion in the financial sector.

**Originality/value:** This paper provides empirical data on Polish consumers' online insurance purchasing behavior. It offers valuable insights into digital trends in the insurance sector and supports strategic decision-making in marketing and service design.

**Keywords:** Digital insurance, consumer behavior, distribution channels,

**Category of the paper:** research paper.

# 1. Introduction

The digital transformation of financial services has profoundly reshaped consumer expectations, access channels, and competitive dynamics across global markets. In this evolving landscape, the insurance sector — traditionally dependent on interpersonal relationships and agent-led distribution — is undergoing a fundamental shift. Digitalisation has moved from being a supplementary tool to becoming a central strategic pillar for insurers seeking to meet growing demands for convenience, price transparency, and user autonomy.

In Poland, this transformation reflects broader European trends, with online platforms, comparison engines, and mobile applications increasingly complementing or replacing conventional agent-based models. Nevertheless, the pace and scope of digital adoption vary significantly across countries and demographic groups, highlighting the importance of local market analyses. While international literature on digital insurance is expanding, the specific behavioral patterns of Polish consumers remain underrepresented. Existing studies tend to generalize findings from Western European markets, neglecting the socio-economic, cultural, and infrastructural nuances that characterize the Polish context.

Moreover, prior research has largely focused on macroeconomic trends or technological innovation, often overlooking micro-level consumer motivations and trust-related concerns — especially relevant in the case of insurance, where perceived risk and the need for personalized support are high. Consequently, there is a clear research gap concerning how Polish consumers choose between digital and traditional insurance distribution channels, what drives their decisions, and what barriers persist in adopting online solutions.

This study aims to fill that gap by providing an original, empirical analysis of online insurance purchasing behavior in Poland. It investigates the frequency and context of digital purchases, explores trust and convenience factors, and identifies the key motivations and concerns shaping channel preferences. The originality of this research lies in its integrated approach, combining theoretical foundations with primary data collected through a structured online survey. In doing so, it reveals the coexistence of digital and traditional models, suggesting that hybrid strategies may offer the most effective path forward for insurers.

The findings not only contribute to the academic discourse on insurance distribution but also provide practical recommendations for companies designing omnichannel systems that respond to diverse consumer needs. By focusing on the Polish market — a rapidly growing yet understudied segment — this article adds important insights into the broader European insurance landscape.

To address the identified research gap, this study poses the following research questions: What factors influence Polish consumers' choice between online and traditional insurance distribution channels? How significant are trust, convenience, and price sensitivity in shaping these choices? What barriers discourage consumers from adopting digital channels? These questions guide the empirical analysis and support the development of practical recommendations for insurers operating in an increasingly hybrid distribution environment.

The remainder of the article is structured as follows: Section 2 presents the literature review on insurance distribution models and digital adoption trends. Section 3 describes the research methodology, including the survey design and respondent profile. Section 4 discusses the empirical results and key patterns identified. Finally, Section 5 outlines conclusions and practical implications, as well as directions for future research.

## 2. Insurance Distribution in Poland: Traditional and Digital Channels

The insurance distribution system functions within the broader context of the financial system, which serves as a mechanism for allocating and transferring purchasing power among economic entities. According to Zabielski (2004), the financial system consists of financial instruments, markets, institutions, and the principles governing their operations. Insurance services, as part of this system, represent a specific type of financial instrument - intangible, contractual, and based on risk redistribution mechanisms. Their provision and sale require tailored distribution models, adapted to the complexity of the service and the expectations of consumers.

The insurance system in Poland reflects the structure of a mixed symbolic and operational system (Kluszczyńska, 2011). It involves public entities (e.g., ZUS) and private market participants such as insurance companies, agents, brokers, and increasingly also technology-based intermediaries like comparison platforms and mobile applications. This dualistic nature - public versus private, traditional versus digital - reflects the broader evolution of distribution strategies in the insurance sector.

Insurance distribution, traditionally carried out through agents, brokers, and branch offices, has evolved under the influence of digitalization. New online channels - such as mobile applications, dedicated customer portals, and e-commerce-style platforms - are increasingly supplementing or even replacing physical intermediaries. As Nowak (2016) note, e-commerce in the insurance market enables cost reduction, faster service, and direct customer engagement, offering added convenience and real-time service personalization.

Despite their advantages, digital channels posed challenges in terms of building trust, reducing perceived risk, and ensuring legal compliance. These concerns were particularly important in insurance, where consumers are often wary of purchasing services without personal consultation. Jończyk in 2010 highlighted the importance of regulatory and technological safeguards to maintain data protection and system transparency - key factors for the long-term sustainability of digital insurance models.

The Polish insurance distribution system has evolved significantly since the liberalisation of the market in the early 1990s, marked by the Insurance Activity Act of 1990. This legislation laid the foundation for the privatisation and demonopolisation of insurance services, aligning the Polish market with European Union standards (Szumlicz, 2005; Wieteski, 2022; Janikowski, 2015). It enabled insurance companies to adopt a multichannel strategy that includes both traditional and modern (online) distribution models.

Traditional distribution channels are primarily composed of sales through agents, brokers, and direct sales via insurers' own branch network. Data from the Polish Financial Supervision Authority confirms this: in 2021, 55.67% of gross written premiums in life insurance (Class I) were generated via agents, with a strong focus on individual insurance products (98.58%). Brokers played a secondary role (3.94%), particularly in group insurance contracts, and direct sales represented 40.33% of gross premiums, driven mainly by group insurance products.

In recent years, direct and online sales channels have become increasingly important in insurance distribution. This includes websites, mobile platforms, and call centers, particularly for simple insurance products such as third-party liability (OC), travel insurance, and personal accident coverage (NNW). The so-called "direct" model allows customers to access policies without the involvement of intermediaries, offering convenience, lower costs, and broad accessibility. The development of online insurance platforms posed a competitive threat to traditional intermediaries. Yet, over time, these direct insurers also began to collaborate with agents to increase sales volumes and scale their distribution (Iwanicz-Drozdowska, 2018).

From a theoretical perspective, the distribution of insurance may be analysed from institutional, functional, and systemic viewpoints. The institutional approach focuses on entities involved in transferring insurance products from insurers to consumers (e.g., agents, brokers, bank branches). The functional approach examines the flow of value and information between participants, while the systemic view considers the entire network of relationships that structure insurance delivery (Szumlicz, 2020).

In recent years, the Polish insurance market has been undergoing a significant digital transformation, reshaping consumer behavior and distribution models. According to Mordor Intelligence (Mordor Intelligence, 2024), the Polish life and non-life insurance market is projected to grow from USD 23.51 billion in 2025 to USD 33.26 billion by 2030, with a compound annual growth rate (CAGR) of 7.19%. A key driver of this expansion is the increasing adoption of digital distribution channels, which are gradually replacing traditional models based on agents and in-person offices. Similarly, the European digital insurance market is witnessing accelerated development, as noted in the Grand View Research report (Grand View Research, 2024), driven by growing internet penetration, rising demand for personalized and on-demand insurance solutions, and the integration of AI-powered platforms in customer service. These trends suggest a fundamental shift in how consumers engage with insurance providers, highlighting the need to further explore the adoption patterns and consumer preferences regarding online insurance channels in the Polish context.

Klapkiv, Szymańska, and Bednarczyk (2018) highlight the dynamic growth of online channels in the distribution of non-life insurance in Poland between 2005 and 2016. During this period, the share of gross written premiums from internet sales in total direct sales (Branch II) increased from a mere 0.01% to 17.35%. Their empirical analysis revealed a strong correlation between the level of household internet penetration and the growth of online insurance sales, supported by an exponential regression model with a high coefficient of determination ($R^2 = 86.84\%$). The authors argue that both technological advancements and legislative changes, such as the implementation of the EU Insurance Distribution Directive (2016/97), have contributed to the increasing importance of digital channels. Furthermore, the growing use of mobile applications, which allow customers to purchase insurance, report claims, and manage their policies, is significantly reshaping consumer behavior and enhancing the competitive edge of insurance providers (Szymańska et al., 2018)

Despite the growing importance of digital channels in the insurance sector, existing literature provides limited insights into the specific purchasing behaviors of insurance consumers in Poland. Given the rapid digital transformation of the insurance industry and the evolving expectations of customers, especially among younger demographics, there is a clear need for further empirical research in this area. Exploring the determinants of insurance purchasing behavior in the Polish context offers valuable opportunities to inform both academic understanding and strategic decision-making in the industry.

## 3. Research methodology and case description

This study was designed to investigate consumer preferences and perceptions regarding different insurance distribution models in Poland, with a particular focus on the growing relevance of online channels. Given the dynamic evolution of digital solutions and shifting consumer behaviors, the research aimed to diagnose the advantages and disadvantages of traditional versus innovative models of insurance distribution.

The research adopted a quantitative diagnostic approach, which allowed the author to reach a broad spectrum of respondents and enabled the statistical generalization of the findings. This methodology was chosen due to its capacity to reveal measurable patterns and relationships between consumer demographics and distribution channel preferences. The central aim of the research was to identify the key benefits and limitations of both traditional (e.g., face-to-face with agents) and digital (e.g., insurer websites, comparison platforms) methods of purchasing insurance products.

The primary research hypothesis stated that online insurance distribution channels are gaining increasing market share in comparison to traditional distribution models. To verify this hypothesis, a structured online survey questionnaire was employed, consisting of both closed-ended and semi-open questions. The instrument was designed to ensure clarity and accessibility for a diverse group of respondents.

The survey covered the following key areas:

- Demographic profile of respondents (e.g., gender, age, education, place of residence).
- Insurance purchasing behavior, including frequency and method of purchasing policies.
- Channel preferences, with respondents evaluating different distribution methods (e.g., agent-based, online direct, brokers).
- Perceptions of key decision-making factors, such as price sensitivity, convenience, and trust.

The research sample included 100 respondents, offering a diverse cross-section of Polish consumers. Analysis focused on identifying statistically relevant trends and correlating consumer profiles with channel preferences.

Respondents in the study represented a diverse demographic profile. The majority were women (67%), while men accounted for 33% of the sample. In terms of age, most participants (55%) were between 26 and 45 years old, with 10% aged 18-25 and 14% over 55. Educationally, the largest group held a bachelor's degree (45%), followed by those with secondary education (30%) and a master's degree or higher (20%), while only 5% had primary education. As for place of residence, the sample included both urban and rural populations: 15% lived in rural areas, 30% in towns up to 100,000 inhabitants, 20% in medium-sized cities (100,000-500,000), and 25% in large cities with over 500,000 residents.

The research technique used was the anonymous online questionnaire, allowing for efficient data collection while ensuring respondent comfort and privacy. The results were interpreted with awareness of possible limitations, including the relatively small sample size and non-probabilistic sampling.

## 4. Research results

A summary of the survey results concerning insurance distribution channel preferences is presented in pie charts numbered 1 through 6. Each chart corresponds to one of the six key survey questions posed to respondents. The data illustrates customers' current choices, expectations, and satisfaction levels regarding various insurance distribution channels, including traditional in-person options and digital solutions.

The charts offer a clear visual representation of customer tendencies, highlighting the most preferred methods of purchasing insurance, reasons behind those preferences, and potential areas for development. Dominant preferences in each area have been emphasized in the chart legends, allowing for quick identification of prevailing trends and facilitating comparative analysis across the surveyed sample.

In the Figure 1 respondents were asked whether they had had experience of purchasing insurance via the internet. The largest number of respondents ticked the answer "more than once", confirming the growing popularity and acceptance of digital channels in the purchase of insurance products. A smaller but still significant group answered "once", indicating their first attempts to use the internet for this purpose. The smallest percentage of respondents indicated "no", indicating no experience with buying insurance online. These data confirm that digital channels are no longer a novelty and have gained the trust of a large proportion of consumers. The regular use of the internet to conclude an insurance contract means that users are not only testing this way of buying, but see it as convenient and secure enough. For insurers, this is a signal that further development and optimisation of digital platforms is not only advisable, but even necessary.

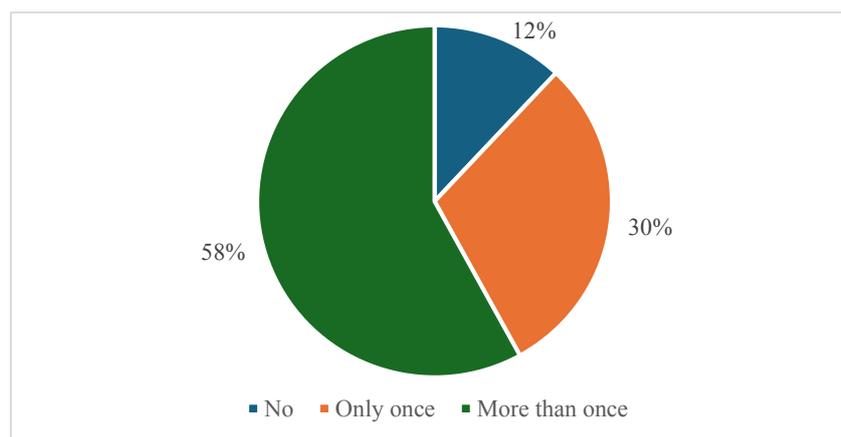

**Figure 1.** Frequency of purchasing insurance online.

In the second graph we see how respondents indicated which insurance distribution channel they use the most. The most popular was the website and in second place the insurance agent and which shows that despite the development of technology, the traditional form of human contact still inspires trust. However, insurance comparison sites also ranked highly, demonstrating that customers value independence, speed and the ability to compare quotes. The insurance distribution market is clearly moving towards a hybrid model. Customers still need the advice of an agent, but they are increasingly using digital channels as a supplement or pre-selection tool for offers. For the industry, this means that traditional and online channels need to be coherently integrated - for example, enabling a customer to start a purchase online and complete it with the help of an advisor.

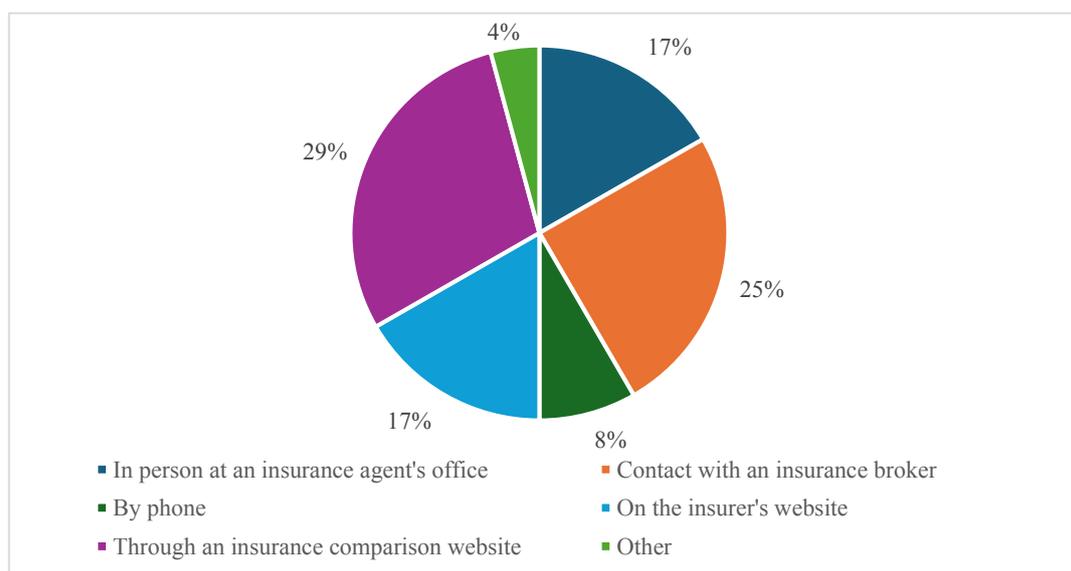

**Figure 2.** Preferred insurance distribution channel.

The third pie chart shows the distribution of respondents' answers to the question on the importance of price when choosing insurance. As many as 70% of respondents consider price to be an important or very important element when choosing a policy purchase channel. This shows that the insurance market in Poland remains highly cost-sensitive, which may be due to comparable offers and low functional differentiation of products (especially third-party liability, accident insurance or travel insurance). The high importance of price is fostered by the growing popularity of digital sales channels, such as insurance comparison sites or mobile apps, which allow customers to quickly collate offers from a price perspective.

22% of those surveyed are not primarily driven by price, showing that there is a group of customers willing to pay more for quality, recommendations or availability of advice. For insurance companies, this is a signal to diversify communication channels and sales arguments - not only through discounts, but also through added values (e.g. fast claims settlement, apps, chat consultants).

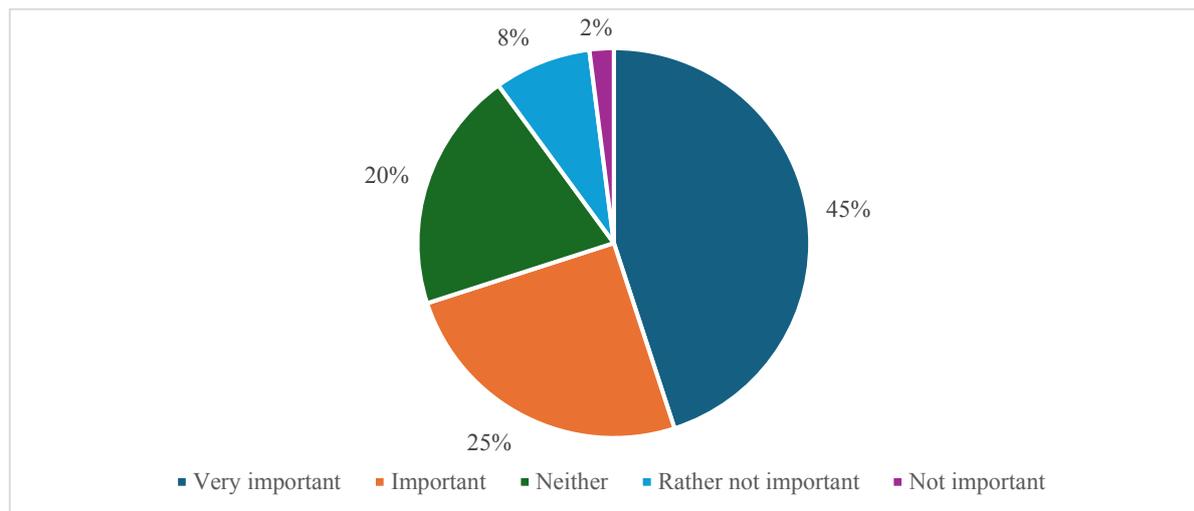

**Figure 3.** Importance of price in channel selection.

In Figure 4 shows the distributions of responses respondents were asked to rate the attractiveness of buying insurance online compared to traditional methods, such as buying through an agent, in a branch or over the phone. As many as 60 per cent of respondents find online insurance purchases more attractive than traditional methods. This is a very strong indication that the market not only accepts digital channels, but actually prefers them - mainly due to the convenience, speed and availability of quotes.

However, it is worth noting that 20% of the respondents do not have a clear opinion and 10% make their decision dependent on the situation. This may be due to the type of insurance (e.g. more complex products like health or property often require contact with an advisor), level of knowledge, or previous purchasing experience. Only 10% found online channels unattractive. For insurers, this is a clear signal that although the majority of customers have converted to the internet, there is still a group that requires additional support and education.

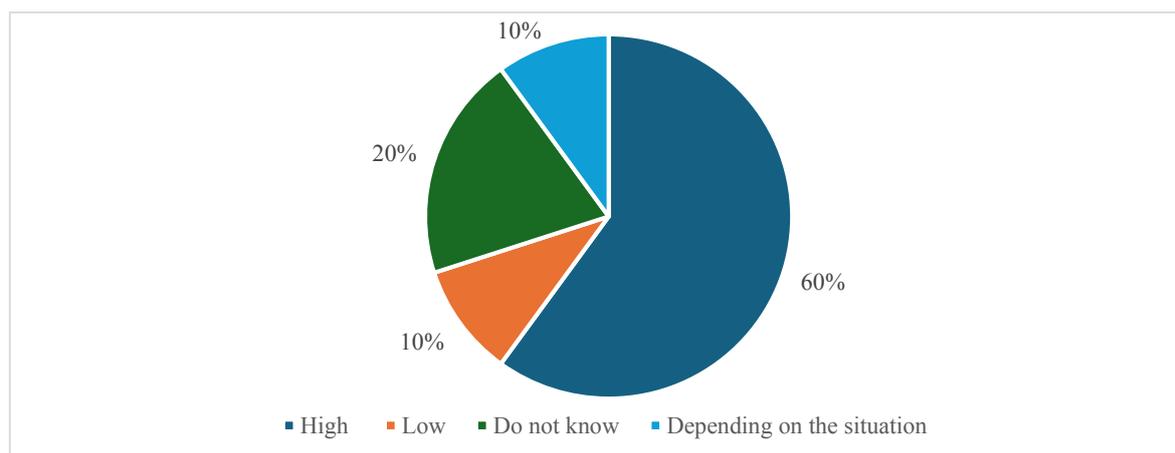

**Figure 4.** Attractiveness of online versus traditional insurance purchasing.

Figure 5 compares the most important factors influencing the choice of insurance distribution channel, taking into account both the advantages of online channels and the benefits of contacting an agent. Respondents indicated various decision-making aspects, which were grouped into three categories: factors related to policy choice, advantages of digital channels and advantages of the agent model. For those choosing a policy, the most important criterion was price, which still dominates the purchase decision process. In the area of digital channel advantages, consumers particularly valued convenience, the availability of services at any time and the possibility of a quick transaction without having to leave home. The ease of comparing offers was also key, showing that users value transparency and control over the purchasing process. In contrast, those using traditional contact with an agent indicated very different values. The most common values emphasised were the importance of personal contact and the opportunity to receive detailed explanations and a personalised approach. Consumers expecting a more comprehensive service and wishing to negotiate insurance terms showed a greater attachment to the agent model. Despite the increasing role of digital channels, trust in traditional forms of sales and the opportunity to speak to an adviser remains important for some customers. The results of the chart confirm the diversity of consumer needs and show that an effective insurance distribution strategy should integrate the advantages of both models - online and in-store - offering both a quick, self-service purchase and the opportunity to consult an expert when needed.

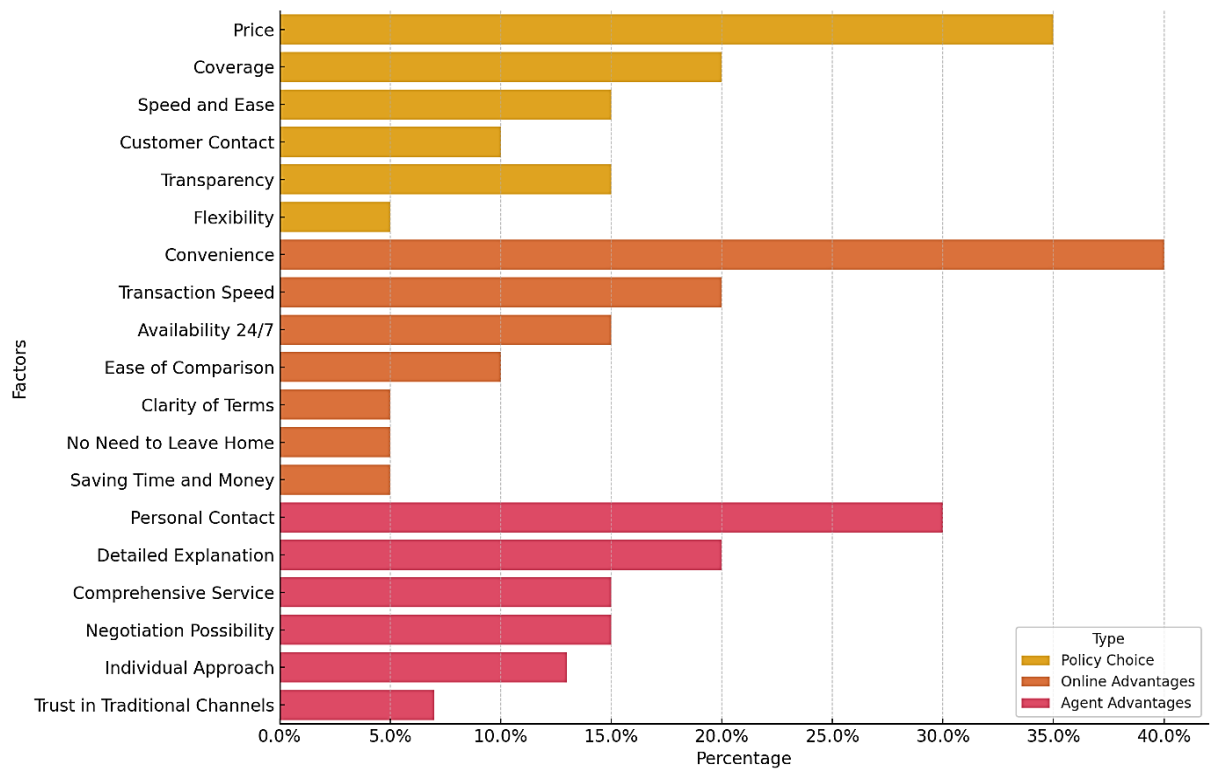

**Figure 5.** Factors influencing distribution channel selection.

The final chart illustrates the factors that discourage consumers from purchasing insurance policies through online channels. Concerns about the security of personal data were the most frequently cited barrier, demonstrating that the issue of privacy and information protection remains central to consumer decisions. Respondents also expressed a lack of confidence in online transactions themselves, which may be a result of general uncertainty towards digital services in the financial sector. The vagueness of policy terms and conditions was also a concern, suggesting that it is important for many people to be able to receive detailed explanations, which are often lacking in automated sales systems. The preference for face-to-face contact with an adviser, which provides a sense of security and the opportunity to ask questions, was also a significant reason for reluctance to buy online. Some respondents also indicated difficulties in understanding the online purchasing process itself, which may be due to the platforms' lack of intuitiveness or users' low digital proficiency. Although online channels are growing in popularity, the chart data shows that many people still perceive them as risky or too complicated. This means that insurers looking to increase the effectiveness of digital sales need to invest not only in technology, but more importantly in customer education, clarity of offer language and better systems security. These responses point to the need for a more empathetic approach to the design of digital sales channels that address not only the technical but also the emotional needs of users.

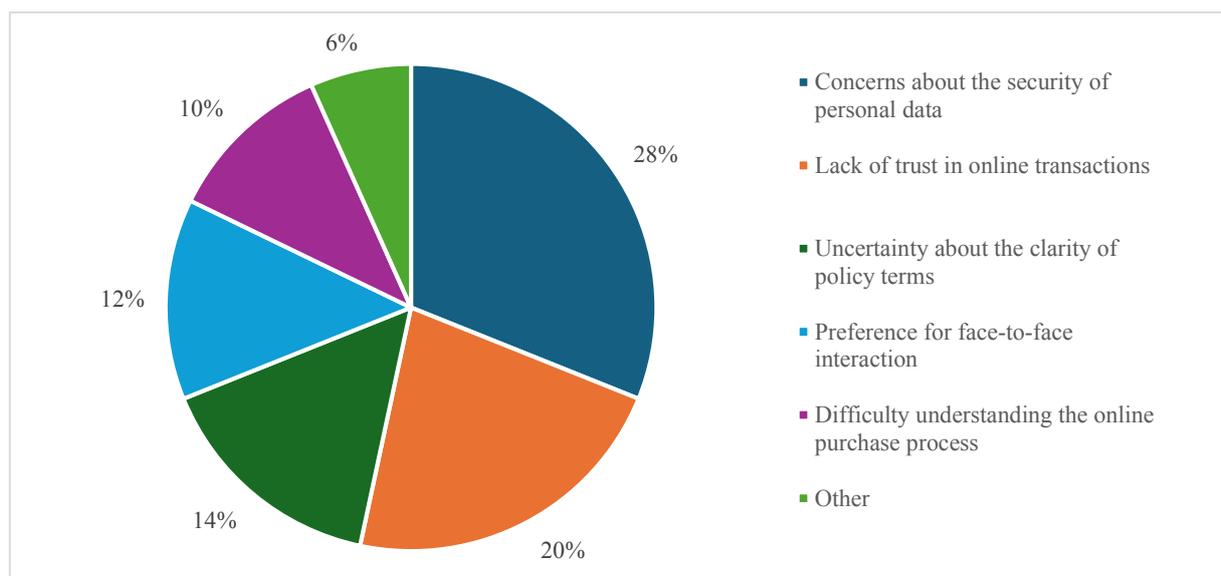

**Figure 6.** Factors discouraging online insurance purchases.

## 5. Conclusions and future research implications

The results clearly show that Polish insurance consumers are actively adopting digital channels, treating them no longer as a novelty, but as a convenient, accessible and fully-fledged tool for purchasing financial services. The majority of respondents have had experience of buying insurance online - often multiple times - which confirms the growing confidence and stabilisation of digital purchasing behaviour in this market segment.

At the same time, the still strong position of insurance agents and the clear need for face-to-face contact testify to the ongoing transformation of the distribution system towards a hybrid model. Customers expect to be able to buy and compare quotes online themselves, but still need expert support in many cases, especially with more complex products. This coincides with earlier analyses by Klapkiv, Szymanska and Bednarczyk (2018), who highlighted the growing importance of online platforms, but at the same time emphasised the need to integrate distribution channels. The results of the current study show that the future of the insurance market does not lie solely in digitalisation, but in the skilful combination of automation with personalisation and empathy.

Price appeared to be the dominant factor influencing the choice of purchase channel, confirming the strong price sensitivity of customers - as indicated by the 2021 FSA data and the conclusions of the Mordor Intelligence report (2024). However, around a quarter of respondents said they were willing to pay a higher price in exchange for better service, recommendations or advisor availability, which provides space for premium offers and value-added differentiation strategies.

The high level of perceived appeal of online shopping shows that for most consumers, digital channels offer not only convenience and speed, but also a sense of control and independence. These findings are in line with earlier observations by Nowak (2016), who emphasised the importance of direct customer engagement and personalisation in e-commerce channels. However, a significant group of respondents still report a lack of trust in online transactions, concerns about the security of personal data and difficulty in understanding the purchase process. These barriers coincide with the analysis of Jończyk (2010), who pointed out the need for transparency and consumer protection in digital services.

The data also reveals clear differences in perceived benefits between online and agency channels. While digital channels dominate in aspects such as convenience, 24/7 availability and ease of comparing quotes, traditional channels are chosen for personal contact, the opportunity to ask questions, negotiate and explain terms and conditions in detail. This shows that purchasing decisions in insurance are not only the result of financial calculation, but also linked to emotions, trust and the need for security.

From a strategic perspective, insurance companies should therefore focus not only on developing digital tools, but also on improving the user experience. It is necessary to design online sales channels in an intuitive, transparent and empathetic way - taking into account users' cognitive and emotional barriers. Enabling technologies such as chatbots, video advisors, interactive calculators or omnichannel platforms that integrate customer data regardless of the contact channel can play a special role here.

In light of the results of this study, several recommendations can also be made for future research. Firstly, it would be worth extending the sample to a larger and more demographically diverse population to get a deeper picture of market segmentation. Secondly, it would be important to explore the relationship between a specific type of insurance and the preferred purchase channel, which could help to better tailor the marketing strategy. Thirdly, it would be useful to look at the impact of artificial intelligence and automation on the perceived quality of service in online channels, which is becoming increasingly important in the context of insurtech and embedded insurance development.

In conclusion, the study confirms that the digitalisation of insurance distribution in Poland is progressing rapidly, but requires further support in the areas of trust, simplicity and consumer education. Companies that can flexibly combine technology with a human approach will have the best chance of successfully reaching different customer segments and maintaining their loyalty in the rapidly changing financial services market environment.